\documentclass[12pt,a4paper]{article}
   \usepackage{epsfig}
   \usepackage[reqno]{amstex}
\normalsize
\advance\textheight by \topskip
\topmargin \paperheight \advance\topmargin by -\textheight
\divide\topmargin by 2 \advance\topmargin by -1in
\advance\topmargin by -\baselineskip
\oddsidemargin \paperwidth \advance\oddsidemargin by -\textwidth
\divide\oddsidemargin by 2 \advance\oddsidemargin by -1in
\evensidemargin \oddsidemargin
\makeatletter
\renewcommand{\@makecaption}[2]{%
  \vskip\abovecaptionskip
  {\small
     \sbox\@tempboxa{#1.\ -- #2}%
     \ifdim \wd\@tempboxa >\hsize
       #1.\ -- #2\par
     \else
       \hbox to\hsize{\hfil\box\@tempboxa\hfil}%
     \fi
  }
  \vskip\belowcaptionskip%
}
\makeatother
\makeatletter
\renewenvironment{thebibliography}[1]%
{%
   \list{\@biblabel{\arabic{enumiv}}}%
   {\settowidth\labelwidth{\@biblabel{#1}}%
   \leftmargin\labelwidth
   \advance\leftmargin\labelsep
	\@openbib@code
   \usecounter{enumiv}%
   \let\p@enumiv\@empty
   }%
   \sloppy\clubpenalty4000\widowpenalty4000%
   \sfcode`\.=\@m%
   \itemsep 0pt plus 1.0pt minus 0pt
}%
{%
   \def\@noitemerr
   {\@latex@warning{Empty `thebibliography' environment}}%
   \endlist%
}
\makeatother
\def\E#1{\left\langle#1\right\rangle }
\def\e{{\rm e}}
\def\d{{\rm d}}

\let\rho=\varrho
\def\minifrac#1#2{{\textstyle\frac{\scriptstyle #1}{\scriptstyle #2}}}
\begin{document}
%--------------------------------------------------------------------
\hbox{}\vskip70pt

\begin{flushleft}\bfseries\Large
  Thermal Equilibrium with the Wiener Potential: Testing the
  Replica Variational Approximation
\end{flushleft}\vskip14pt

\newlength{\dummylength}\dummylength\textwidth \advance\dummylength by -10mm
\begin{flushright}
\begin{minipage}[t]{\dummylength}
   \textsc{K. Broderix} and \textsc{R. Kree} \\[7pt]
\small{\itshape
   Institut f\"ur Theoretische Physik, Georg-August-Universit\"at\\
   Bunsenstr. 9, D-37073 G\"ottingen, Germany\\
   E-mail:
   kree{\makeatletter{}@\makeatother}theorie.physik.uni-goettingen.de
   \\[14pt]
}
   (Version of Juli 24, 1995)\\[12pt]
  {\bfseries Abstract.} --
   We consider the statistical mechanics of a classical
   particle in a one-dimensional box subjected to a random potential
   which constitutes a Wiener process on the coordinate axis.  The
   distribution of the free energy and all correlation functions of the
   Gibbs states may be calculated exactly as a function of the box
   length and temperature. This allows for a detailed test of results
   obtained by the replica variational approximation scheme. We show
   that this scheme provides a reasonable estimate of the averaged free
   energy. Furthermore our results shed more light on the validity of
   the concept of approximate ultrametricity which is a central
   assumption of the replica variational method.
\end{minipage}
\end{flushright}\vskip28pt\par
The replica trick and Parisi's scheme of replica symmetry breaking
within mean field theories provided a promising step towards the
application of powerful methods of field theory to systems with
quenched random couplings \cite{MePa87}. More recently these concepts
have been applied to a variety of physical systems as an approximation
scheme within a framework called ``replica variational approximation''
(RVA). The strategy of the RVA \cite{MePa9091} is to calculate
variational bounds on the moments of the canonical partition sum
within an ensemble of quenched disordered couplings. For integer $n>0$
these bounds can be achieved from standard inequalities of equilibrium
statistical mechanics.  Disorder averaged physical quantities are
estimated from these bounds by using the ``$n\to 0$'' trick
$\ln(x)=\lim_{n\to 0}(x^n-1)/n$.  The method has been tested using toy
models of manifolds in random media \cite{MePa92,Eng93}.  In the limit
of a manifold embedded in infinite dimensional space the RVA with a
Gaussian trial distribution becomes exact. For a zero dimensional
manifold in one dimensional space the test has used exact results on
certain linear combinations of moments of the manifold's position
which have been obtained some time ago \cite{ScVi88}. In a more recent
work \cite{Opp93} the probability distribution of the free energy of a
related model defined on the half axis has also been obtained.

The purpose of the present letter is twofold. First, we give a
complete solution of the zero dimensional toy model.  We use the
expression ``complete'' to indicate that the solution does not only
provide an explicit analytical expression for the probability density
of the free energy for arbitrary temperatures and system sizes but
also allows us to calculate all correlation functions of the Gibbs
state (canonical distribution). In this way we completely characterize
the \emph{probability functional} of the Gibbs state which contains
all the statistical information on the thermostatics of the disordered
system. Second we use our results to test the quality of the RVA.

We consider the position $x$ of a classical particle in a
one-dimensional box $0\leq x \leq L$ subjected to a random potential.
On the one hand we will use for the random potential the standard
Wiener process $W(x)$, that is, the Gaussian Markov process with
averages $\E{W(x)}=0$ and $\E{W(x)W(y)}=\min(x,y)$.  But mainly we
will focus our interest on its variant $V(x):=W(x)-W(L)/2$.  Below we
will refer to these potentials as the $W$- and $V$-ensemble,
respectively.  A single particle moving under the influence of the
potential $V(x)$ may be considered as a continuum model of a kink in
an Ising chain with random fields uncorrelated at different sites
\cite{ViSe83} whereas the $W$-ensemble is related to the asymptotic
behaviour of a directed polymer in a random potential in $1+1$
dimensions \cite{Par90}.  Note that the Gaussian random potential
$V(x)$ is equivalently characterized by the translationally invariant
averages $\E{V(x)}=0$ and $\E{V(x)V(y)}=(L/2-|x-y|)/2$. The position
$x$ of a particle in contact with a heat bath of fixed inverse
temperature $\beta$ is distributed according to the Gibbs state
$\Omega_{\beta,L}(x):=\exp\{-\beta V(x)\}/Z_{\beta}(L)$. In the
following we will be interested in statistical properties of this
state, the corresponding partition sum $Z_{\beta}(L):=\int_{0}^{L}\d
x\,\exp\{-\beta V(x)\}$ and the free energy
$F_\beta(L):=-\beta^{-1}\ln Z_\beta(L)$ within the $V$-ensemble. Our
general strategy will be to calculate such properties from
corresponding properties of the $W$-ensemble with partition sum
$\hat{Z}_{\beta}(L):=\int_{0}^{L}\d x\, \exp\{-\beta W(x)\}$ and free
energy $\hat{F}_\beta(L):=-\beta^{-1}\ln \hat{Z}_\beta(L)$.

Throughout our calculations we find it convenient to use scaled
variables for locations, potentials and partition sums defined as:
$s:=\beta^{2}x/8$, $v(s):=-(\beta/2)V(8s/\beta^{2})$,
$w(s):=-(\beta/2)W(8s/\beta^{2})$,
$z(s):=(\beta^{2}/2)Z_{\beta}(8s/\beta^{2})$, and
$\hat{z}(s):=(\beta^{2}/2)\hat{Z}_{\beta}(8s/\beta^{2})$.  The scaled
system length will be denoted by $l:=\beta^{2}L/8$.  The transition
density $p_s(w,\hat{z}|w_0,\hat{z}_0)$ of the homogeneous Markov
process $(w(s),\hat{z}(s))$ may be determined from the Fokker-Planck
equation
\begin{equation} \label{FP}
   \partial_{s} p_s(w,\hat{z}|w_0,\hat{z}_0)
   = \left( \partial_{w}^{2}-4\e^{2w}\partial_{\hat{z}} \right)
   p_s(w,\hat{z}|w_0,\hat{z}_0)
\end{equation}
with initial condition $p_0(w,\hat{z}|w_0,\hat{z}_0) =
\delta(w-w_0)\,\delta(z-z_0)$.  This equation is equivalent to the
Langevin equations $\d w(s)/\d s = \sqrt{2}\,\xi(s)$ and
$\d\hat{z}(s)/\d s = 4\e^{2w(s)}$ following immediately from the
definitions of $w(s)$ and $\hat{z}(s)$. Here $\xi$ denotes standard
Gaussian white noise with $\E{\xi(s)}=0$ and
$\E{\xi(s)\xi(s')}=\delta(s-s')$.  To obtain the joint probability
density of $v(s)$ and $z(s)$ from the transition density of $w(s)$ and
$\hat{z}(s)$ we use the relation
\begin{equation}\label{TRAFO}\begin{split}
  & \Big\langle\delta(z(s)-z)\,\delta(v(s)-v)\Big\rangle
\\ \hbox{}=\hbox{}&
\int_{-\infty}^\infty\d w\, \e^w \int_0^\infty\d\hat{z}\,
p_{l-s}\!\left( w,\hat{z} \left| v+\minifrac{1}{2}w,\e^w z
\right.\right) p_s\!\left(\left.  v+\minifrac{1}{2}w,\e^w z \right|
0,0 \right).
\end{split}\end{equation}
Note that according to (\ref{FP}) one has
\begin{equation} \label{TRANS}
  p_s(w,\hat{z}|w_0,\hat{z}_0) = p_s(w,\hat{z}-\hat{z}_0|w_0,0).
\end{equation}

To solve the Fokker-Planck equation (\ref{FP}) we perform a Laplace
transform with respect to $\hat{z}$ and introduce
\begin{equation} \label{LT}
  \tilde{p}_s(w,\rho|w_0) := \int_{0}^{\infty}\d \hat{z}\,
  \e^{-\hat{z}\rho} p_s(w,\hat{z}|w_0,0)
\end{equation}
which obeys the equation
\begin{equation} \label{LTFP}
  \left(\partial_{s}-\partial_{w}^{2}+4\rho\,\e^{2w}\right)
    \tilde{p}_s(w,\rho|w_0) = 0
\end{equation}
with initial condition $\tilde{p}_0(w,\rho|w_0) = \delta(w-w_0)$.
Thus we see that $\tilde{p}$ becomes a Green function of a Bessel-type
differential equation. It may be given in terms of an eigenfunction
expansion
\begin{equation} \label{SOL}
  \tilde{p}_s(w,\rho|w_0) = \frac{2}{\pi^2} \int_{0}^{\infty}\d\nu \,
  \nu\sinh(\pi\nu)\,\e^{-\nu^2 s}\, K_{i\nu}\!\left(2\sqrt{\rho}\,\e^w
\right) K_{i\nu}\!\left(2\sqrt{\rho}\,\e^{w_0} \right),
\end{equation}
where $K_{i\nu}(y)=\int_{0}^{\infty}\d t\, \cos(t\nu)\, \exp\{-y\cosh
t\}$ denotes the modified Bessel function of third kind with index
$i\nu$, see \cite[Chap.~9]{AbSt72}.  As to the validity of (\ref{SOL})
we remark that
$\e^{-\nu^2s}\,K_{i\nu}\!\left(2\sqrt{\rho}\,\e^w\right)$ obeys
(\ref{LTFP}) for all $\nu$ which can be checked with the help of the
differential equation \cite[Eq.~9.6.1]{AbSt72} for the modified Bessel
functions. Furthermore, the right-hand side of (\ref{SOL}) reproduces
the demanded initial condition which can be inferred from the pair
\cite[p.~173]{Erd54} of reciprocal formulas for the
Kontorovich-Lebedev transformation.

In the sequel we will summarize some of our results on the statistics
of the free energy and the Gibbs state.  All of these results can be
obtained from the fundamental solution (\ref{SOL}). A detailed
exposition of the calculations and the results will be given in
\cite{BrKr??}.

For the discussion of the statistical properties of the free energies it
is convenient to introduce their scaled variants
$f(l):=-l^{-1/2}\ln(z(l))$ and $\hat{f}(l):=-l^{-1/2}\ln(\hat{z}(l))$
from which the original quantities can be recovered according to
\begin{equation}\label{FUNSCALE}
   F_\beta(L) = \frac{1}{2} \sqrt{\frac{L}{2}}\,
   f\!\left(\frac{\beta^2}{8}L\right)
   + \frac{1}{\beta} \ln\frac{\beta^2}{2}
\end{equation}
and analogously for $\hat{F}_\beta(L)$. As
$z(l)=\e^{-w(l)}\hat{z}(l)$, the averaged free energy is the same for
both ensembles, that is, $\langle f(l)\rangle
=\langle\hat{f}(l)\rangle$. Fluctuations of the free energy, however,
will differ between the two ensembles. For example, one has
$\langle(f(l))^2\rangle=\langle(\hat{f}(l))^2\rangle-2$ as shown in
\cite{BrKr??}.

The density of the free energy $f(l)$ turns out to be
\begin{equation}\label{FDIST}\begin{split}
   \E{\delta(f(l)-f)} = \hbox{}&
   \frac{2}{\pi\sqrt\pi}\, \exp\!\left\{
      \frac{\pi^2}{4l} + f\sqrt l
   \right\} K_0\!\left(2\,\e^{f\sqrt l}\right)
\\ & \times
   \int_0^\infty\d t \, \sinh(t) \, \sin\!\left(\frac{\pi}{2l}\,t\right)
   \exp\!\left\{
       -\frac{t^2}{4l} - 2\, \e^{f\sqrt l}\, \cosh(t)
   \right\}.
\end{split}\end{equation}
In Fig.~\ref{FIG1} a plot of the centered density of $f(l)$ is given
for different values of $l$. An expression analogous to (\ref{FDIST})
for the density of $\hat{f}(l)$ will be given in \cite{BrKr??}.

Of course, the densities allow for a calculation of the averaged free
energies and their fluctuations. Here we only give the asymptotic
expansion of the averaged free energy for $l\to\infty$
\begin{equation}
   \langle f(l)\rangle =
   \langle\hat{f}(l)\rangle =
   -\frac{4}{\sqrt\pi} - \frac{\gamma}{\sqrt l} +
   \frac{\pi\sqrt\pi}{6l}
   + O\!\left(l^{-2}\right),
\end{equation}
where $\gamma=0.5772\dots$ denotes the Euler-Mascheroni constant
\cite[Eq.~6.1.3]{AbSt72}.

%----------------------------------------------------------------------
\begin{figure}[b]
   \vskip-5mm
   \centering\epsfig{file=fig1.eps, width=130mm}
   \caption[0]{%
      Centered density of the scaled free energy $f(l)$ of the
      $V$-ensemble for different system sizes $l=1/10,1,10,100$.%
   }\label{FIG1}
\end{figure}
%----------------------------------------------------------------------

In order to get along with the calculation of the averaged free energy
within the RVA using the standard Gaussian trial distributions we
extend the system to the whole real axis by softening the walls. More
precisely, we use the potential $v(s)+(1-2s/l)^{2M}$, $s$ real,
$M=1,2,3,\dots$, instead of $v(s)$ for the calculations and eventually
perform the limit $M\to\infty$. For this setting we have re-done the
RVA along the lines of \cite{MePa9091}. The details will be given in
\cite{BrKr??}. The results of the RVA to the averaged free energy
$\langle f(l)\rangle\approx f^{\text{RVA}}(l)$ are as follows. For
$l<\sqrt{3\pi}$ the saddle-point equations have a replica symmetric
solution only, resulting in
\begin{equation}
   f^{\text{RVA}}(l) :=
   f^{\text{RS}}(l) :=
   - \frac{1+\sqrt 2}{\sqrt\pi} \nu_l \sqrt l
   - \frac{1}{2\sqrt l}
   \left( 1+\ln\!\left(8\pi l^2\nu_l^2\right)\right),
   \quad\text{for $l<\sqrt{3\pi}$,}
\end{equation}
where $0\le\nu_l\le 1$ solves
$\frac{2l}{\sqrt\pi}\nu_l^3+\nu_l^2=1$. For $l>\sqrt{3\pi}$ a replica
symmetry broken solution exists, too. Since this solution is probably
the stable one, we set
\begin{equation}
   f^{\text{RVA}}(l) :=
   f^{\text{RSB}}(l) :=
   - \frac{3(1+\sqrt 2)}{(12\pi)^{1/4}}
   - \frac{1}{2\sqrt l}
   \left( 3+\sqrt 2 - \ln\!\left(8\pi^2\right)\right),
   \quad\text{for $l>\sqrt{3\pi}$.}
\end{equation}

In Fig.~2 the exact and approximate averaged free energy are compared.
One finds that the RVA constitutes a \emph{lower} bound. This has also
been observed in comparison with simulations \cite{MePa92,Eng93}. To
our knowledge, this phenomenon has not yet been satisfactorily
explained.

%----------------------------------------------------------------------
\begin{figure}[b]
   \vskip5mm
   \centering\epsfig{file=fig2.eps, width=120mm}
   \caption[0]{%
     The main plot shows the RVA $F^{\text{RVA}}_\beta(L)$ to the
     averaged free energy $\langle F_\beta(L)\rangle$ for $L=1$ as a
     function of temperature $1/\beta$. The vertical dashed line
     indicates the temperature, below which replica symmetry breaking
     sets in. For $1/\beta$ near $0$ both curves are dominated by the
     logarithmic contribution in (\ref{FUNSCALE}). Therefore the
     inlets show the scaled variants $f^{\text{RS}}(l)$,
     $f^{\text{RSB}}(l)$ and $\langle f(l)\rangle$, respectively, as a
     function of $1/l$. Although the leading correction of the RVA to
     its ground-state energy for $l\to\infty$ is of correct order
     $l^{-1/2}$, it is clearly seen that it has the wrong sign.
   }\label{FIG2}
\end{figure}
%----------------------------------------------------------------------

Now we turn to the calculation and discussion of the statistical
properties of the Gibbs states. As can be seen from its definition,
$\Omega_{\beta,L}(x)$ is invariant under a change from the $V$- to the
$W$-ensemble. Therefore we may calculate all the correlation functions
of the Gibbs state directly within the $W$-ensemble. The $J$-point
correlation function of the scaled Gibbs state
$\omega_l(s):=(8/\beta^2)\,\Omega_{\beta,8l/\beta^{2}}(8s/\beta^{2})$
can be expressed due to (\ref{TRANS}) in terms of the Lapace
transformed transition density $\tilde{p}$ as
\begin{equation}
   \E{\prod_{j=1}^J\omega_l(s_j)} =
   \frac{4^J}{(J-1)!}\int_0^\infty\d\rho\,\rho^{J-1}\!
   \int_{-\infty}^\infty\!\!\d w_1 \cdots \d w_{J+1}
   \prod_{j=0}^{J} \tilde{p}_{s_{j+1}-s_{j}}(w_{j+1},\rho|w_{j})
   \,\e^{2w_{j}},
\end{equation}
where $0=:s_0\le s_1 \le s_2 \le \cdots s_J \le s_{J+1} := l$ and
$w_0:=0$ is assumed.  With the help of (\ref{SOL}) this expression may
be simplified further.  Eventually we end up with a representation of
the $J$-point correlation function in terms of a $(J+1)$-fold integral
of elementary transcendental functions, see \cite{BrKr??}.

Here we specialize to the most interesting situation of a large system
$l\gg 1$ and large separations between the intermediate points $s_j$.
As to be expected, the leading behaviour
\begin{equation}
   \lim_{l\to\infty}l^J\E{\prod_{j=1}^J\omega_l(l\sigma_j)} =
   \frac{1}{\pi\sqrt{\sigma_1(1-\sigma_1)}} \prod_{j=2}^J
   \delta(\sigma_j-\sigma_1), \quad 0<\sigma_1,\dots,\sigma_J<1,
\end{equation}
of the correlation function reflects the fact that the system
possesses a unique ground state with probability $1$. For
$0<\sigma_1<\sigma_2<\dots<\sigma_J<1$ the next-to-leading correction
turns out to be
\begin{equation}
   \lim_{l\to\infty}l^{(3J-1)/2}\E{\prod_{j=1}^J\omega_l(l\sigma_j)}
   =
   \frac{1}{\pi\sqrt{\sigma_1(1-\sigma_J)}} \,
   \frac{(16\pi)^{(1-J)/2}} {(J-1)!}
   \prod_{j=2}^J (\sigma_j-\sigma_{j-1})^{-3/2}.
\end{equation}
A short distance expansion given in \cite{BrKr??} reveals that the
correlation functions are analytic for small $s_j-s_{j-1}$ and finite
$l$.

These results allow for a discussion of the validity of the RVA on a
deeper level then does (\ref{FDIST}) because we may now check the
complete structure of the \emph{probability functional} of Gibbs
states. Note that the average $\E{\omega(s)}$ will always be a
Gaussian within the replica variational ansatz used in \cite{MePa9091}
whereas it is actually $(\pi\sqrt{s(l-s)})^{-1}$ for large systems.
One might argue that this is far away from a Gaussian. However, the
Gaussian form is inevitably encoded in the ansatz and thus we do not
consider this a serious defect. Much more interesting is a comparison
between the statistics of typical distances $\Delta=s-s'$. This has
been obtained in \cite{MePa9091} within the RVA. The result implies
that the probability density $P(\Delta)$ should decrease like a
Gaussian for large $\Delta$. The exact result can be obtained from
$\E{\omega(s)\omega(s')}$ and is given by
$P(\Delta)\sim(1/\sqrt{16\pi})|\Delta|^{-3/2}$ for large but finite
$l$ and $\Delta$. Thus the RVA severely underestimates the range of
correlations. However, the most crucial feature of the RVA, which is
the hierarchical structure of the ansatz, seems to be in accordance
with a real property of the system, at least in a qualitative sense as
we will now explain.  The geometry of typical distances may be studied
by considering higher order correlations.  For example, consider 3
points $0<s_{1}<s_{2}<s_{3}<l$ with corresponding distances
$\Delta_{12}, \Delta_{13}$ and $\Delta_{23}$.  If we assume $l$ and
all distances to be large but finite, their joint probability density
is given by $P(\Delta_{12},\Delta_{13},
\Delta_{23})\sim(1/32\pi)(\Delta_{12}\Delta_{23})^{-3/2}
\delta(\Delta_{12}+\Delta_{23}-\Delta_{13})$. Thus for fixed $s_{1}$
and $s_{3}$ the intermediate location $s_{2}$ will most probably be
near to one of these points. If we regard these three points as a
``one-dimensional triangle'' we conclude that isoscele triangles with
a shorter third side are statistically preferred. In this sense, there
is a weak, statistical form of ultrametricity present in the system.
An ultrametric geometry also appears as a consequence of the
hierarchical ansatz for the replica symmetry breaking scheme
\cite{MePa87,MePa9091} which thus captures an important feature of the
probability functional of Gibbs states.
\vskip28pt\par

\noindent REFERENCES

%--------------------------------------------------------------------
\end{document}